\documentclass[twocolumn]{aastex62}

\usepackage{graphicx,color} 
\usepackage{hyphenat} 
\usepackage{amsmath, amssymb, amsfonts,amsbsy,amsthm} 

\newcommand{\hatbf}[1]{\mathbf{\hat{#1}}}
\newcommand{\dd}{\mathrm{d}}
\newcommand{\overbar}[1]{\mkern 1.5mu\overline{\mkern-1.5mu#1\mkern-1.5mu}\mkern 1.5mu}

\begin{document}
\title{Gravitational-Wave Geodesy: A New Tool for Validating Detection of the Stochastic Gravitational-Wave Background}
\author{T. A. Callister}
\email{tcallist@caltech.edu}
\author{M. W. Coughlin}
\author{J. B. Kanner}
\affiliation{LIGO Laboratory, California Institute of Technology, Pasadena, CA 91125, USA}
\date{\today}

\begin{abstract}
A valuable target for advanced gravitational-wave detectors is the stochastic gravitational-wave background.
The stochastic background imparts a weak correlated signal into networks of gravitational-wave detectors, and so
standard searches for the gravitational-wave background rely on measuring cross-correlations between pairs of widely-separated detectors.
Stochastic searches, however, can be affected by any other correlated effects which may also be present, including correlated frequency combs and magnetic Schumann resonances.
As stochastic searches become sensitive to ever-weaker signals, it is increasingly important to develop methods to separate a true astrophysical signal from other spurious and/or terrestrial signals.
Here, we describe a novel method to achieve this goal -- gravitational-wave geodesy.
Just as radio geodesy allows for the localization of radio telescopes, so too can observations of the gravitational-wave background be used to infer the positions and orientations of gravitational-wave detectors.
By demanding that a true observation of the gravitational-wave background yield constraints consistent with the baseline's known geometry, we demonstrate that we can successfully validate true observations of the gravitational-wave background while rejecting spurious signals due to correlated terrestrial effects.
\vspace{1cm}
\end{abstract}


\section{Introduction}

The recent Advanced LIGO-Virgo observations of binary black hole \citep{O1BBH,GW170104,GW170608,GW170814} and binary neutron star \citep{GW170817} mergers suggest that the astrophysical stochastic gravitational-wave background may soon be within reach \citep{Implications150914,Implications170817,IsotropicO1,DirectionalO1}.
As the superposition of all gravitational-wave signals too weak to individually detect, the stochastic gravitational-wave background is expected to be dominated by compact binary mergers at cosmological distances \citep{Regimbau2008,Rosado2011,Zhu2011,Wu2012,Zhu2013,Callister2016}.
Although the stochastic background is orders of magnitude weaker than instrumental detector noise, it will nevertheless impart a weak \textit{correlated} signal to pairs of gravitational-wave detectors.
The stochastic background may therefore be detected in the form of excess correlations between widely-separated gravitational-wave detectors \citep{Christensen1992,Allen1999,Romano2017}.


Cross-correlation searches for the stochastic background rely on the assumption that, in the absence of a gravitational-wave signal, the outputs of different gravitational-wave detectors are fundamentally uncorrelated.
The LIGO-Hanford and LIGO-Livingston detectors, for instance, are separated by 3000\,km, with a light travel time of $\approx$\,0.01\,s between sites.
One might therefore reasonably expect them to be safely uncorrelated at $\sim \mathcal{O}(100\,\mathrm{Hz})$, in the frequency band of interest for ground-based detectors.

In reality, however, terrestrial gravitational-wave detectors are \textit{not} truly uncorrelated.
Hanford-Livingston coherence spectra consistently show correlated features that, if not properly identified and removed, can severely contaminate searches for the stochastic gravitational-wave background \citep{Covas2018}.
Schumann resonances are one expected source of terrestrial correlation \citep{Schumann1,Schumann2}.
Global electromagnetic excitations in the cavity formed by the Earth and its ionosphere, Schumann resonances may magnetically couple to Advanced LIGO and Advanced Virgo's test mass suspensions and induce a correlated signal between detectors \citep{Christensen1992,Thrane2013,Thrane2014,Coughlin2016,Coughlin2018}.
Another expected source of correlation is the joint synchronization of electronics at each detector to Global Positioning System (GPS) time.
In Advanced LIGO's O1 observing run, for instance, a strongly-correlated 1\,Hz comb was traced to blinking LED indicators on timing systems independently synchronized to GPS \citep{Covas2018}.


Any undiagnosed terrestrial correlations may yield a false-positive detection of the stochastic gravitational-wave background.
While Schumann resonances and frequency combs represent two known classes of correlation, there may exist others.
The validation of any apparent observation of the stochastic background will therefore require us to answer the following question:
\textit{How likely is an observed correlated signal to be of astrophysical origin, rather than a yet-unidentified source of terrestrial correlation?}

We currently lack the tools to quantitatively answer this question.
Searches for gravitational-wave transients can address this issue through the use of time-slides: the artificial time-shifting of data from one detector relative to another's.
This process eliminates any coherent gravitational-wave signals while preserving all other properties of the data, allowing for accurate estimation of the rate of false positive detections.
In cross-correlation searches for the stochastic background, however, time-slides would not only remove a gravitational-wave signal but also any correlated terrestrial contamination as well.
Time-slides are therefore of limited use in searches for the gravitational-wave background.

Using techniques borrowed from the field of radio geodesy, here we develop a novel method to evaluate the astrophysical significance of an apparent correlated stochastic signal.
Just as interferometric observations of the radio sky can serve to precisely localize radio telescopes on the Earth, we demonstrate that measurements of the gravitational-wave background can be similarly reverse-engineered to infer the separations and relative orientations of gravitational-wave detectors.
By demanding that a true gravitational-wave background yield results consistent with the \textit{known} geometry of our detectors, we can separate true gravitational-wave signals from spurious terrestrial correlations.

First, in Sect. \ref{formalism}, we review search methods for the stochastic gravitational-wave background and introduce gravitational-wave geodesy.
In Sect. \ref{modelSelection}, we use geodesy as the basis of a Bayesian test with which to reject non-astrophysical signals, and in Sect. \ref{demonstration}, we demonstrate this procedure using simulated measurements of both a gravitational-wave background and terrestrial sources of correlation.
Finally, in Sect. \ref{complications}, we discuss potential complications and outline directions for future work.
	
\section{Gravitational-Wave Geodesy}
\label{formalism}

\begin{figure}
\centering
\includegraphics[width=0.48\textwidth]{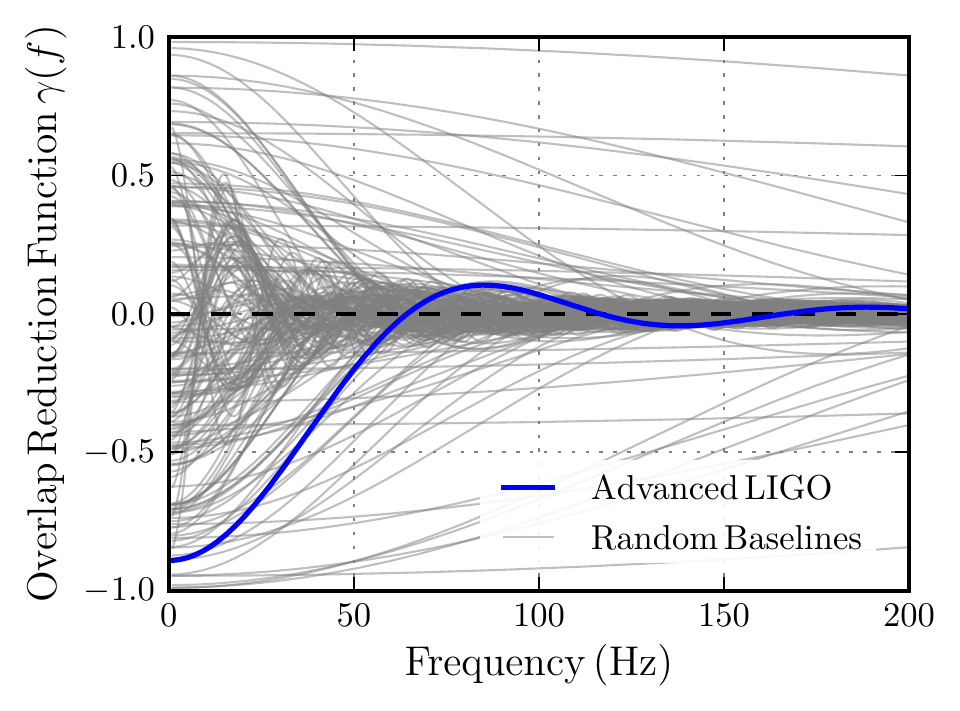}
\caption{
The overlap reduction function $\gamma(f)$ (blue) for the Advanced LIGO's Hanford-Livingston detector baseline.
Alternative baseline geometries have different overlap reduction functions as illustrated by the collection of grey curves, which show overlap reduction functions between hypothetical detectors randomly positioned on Earth's surface.
}
\label{orfFigure}
\end{figure}

The stochastic background is typically described via its energy-density spectrum $\Omega(f)$, defined as the energy density $d\rho_\textsc{gw}$ of gravitational waves per logarithmic frequency interval $d\ln f$ \citep{Allen1999,Romano2017}:
	\begin{equation}
	\label{omega}
	\Omega(f) = \frac{1}{\rho_c} \frac{d\rho_\textsc{gw}}{d\ln f}.
	\end{equation}
The energy-density spectrum is made dimensionless by dividing by the Universe's closure energy density $\rho_c = 3 H_0^2 c^2/(8\pi G)$, where $H_0$ is the Hubble constant, $c$ is the speed of light, and $G$ is Newton's constant.

Searches for the stochastic background seek to measure $\Omega(f)$ by computing the cross-correlation spectrum $\hat C(f)$ between pairs of gravitational-wave detectors:
	\begin{equation}
	\label{crosscorr}
	\hat C(f) = \frac{1}{\Delta T} \frac{20\pi^2}{3 H_0^2} f^3 \,\mathrm{Re} \,\left[\tilde s_1^*(f) \tilde s_2(f)\right],
	\end{equation}
where $\Delta T$ is the time duration of data analyzed and $\tilde s_I(f)$ is the (Fourier domain) strain measured by detector $I$.
Equation \eqref{crosscorr} is normalized such that, for Advanced LIGO, the expectation value of $\hat C(f)$ is \citep{Allen1999}
	\begin{equation}
	\label{avg}
	\langle \hat C(f) \rangle = \gamma(f) \Omega(f).
	\end{equation}
In the weak signal limit, the variance of $\hat C(f)$ is given by $\langle \hat C(f) \hat C(f')\rangle = \delta(f-f') \sigma^2(f)$, with
	\begin{equation}
	\sigma^2(f) = \frac{1}{\Delta T} \left(\frac{10\pi^2}{3 H_0^2}\right)^2 f^6 P_1(f) P_2(f),
	\label{sigma}
	\end{equation}
where $P_i(f)$ is the one-sided noise power spectral density of detector $i$.
Given a model $C_\mathcal{H}(f)$ for the energy-density spectrum of the background, the signal-to-noise ratio (SNR) of a stochastic measurement $\hat C(f)$ is given by the inner product $\mathrm{SNR}^2 = \bigl(\hat C | C_\mathcal{H}\bigr)$, where 
	\begin{equation}
	\label{innerProduct}
	\left(A | B\right) = 2 \int_{0}^\infty \frac{A^*(f) B(f)}{\sigma^2(f)} \mathrm{d}f.
	\end{equation}
	
\begin{figure*}
\centering
\includegraphics[width=\textwidth]{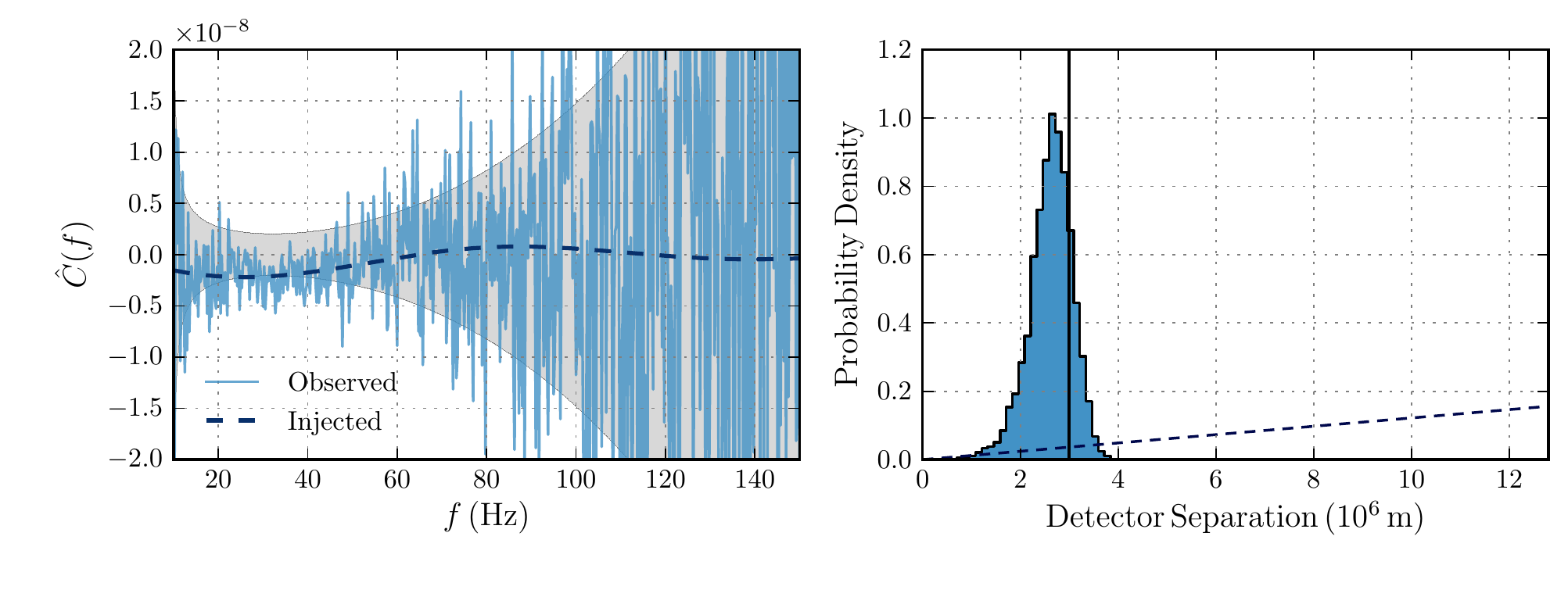}
\caption{
\textit{Left}: Simulated Advanced LIGO cross-correlation measurements (blue) following a three-year observation of an isotropic stochastic gravitational-wave background.
The injected background has energy-density $\Omega(f) = 3.33\times10^{-9} \left(f/25\,\mathrm{Hz}\right)^{2/3}$, corresponding to an expected signal-to-noise ratio of 10 after three years of observation.
The dashed curve shows the expected cross-correlation in the absence of measurement noise, and the grey band indicates $\pm1\sigma$ uncertainties.
\textit{Right}: The posterior on the distance between the LIGO Hanford and Livingston detectors, obtained using the simulated cross-correlation measurements shown on the left.
The dashed line indicates the distance prior used and the vertical black line marks the true Hanford-Livingston separation.
Using the gravitational-wave sky, we self-consistently recover a posterior compatible with the true distance between detectors.
Details regarding parameter estimation are explained in Sect. \ref{modelSelection} below.
}
\label{example}
\end{figure*}
	
The factor $\gamma(f)$ appearing in Eq. \eqref{avg}, called the normalized overlap reduction function, encodes the dependence of the measured correlations on the detector baseline geometry -- the detectors' locations and relative orientations \citep{Christensen1992}.
Advanced LIGO's normalized overlap reduction function is given by
	\begin{equation}
	\label{orf}
	\gamma(f) = \frac{5}{8\pi} \sum_A \int_\mathrm{Sky} F^A_1(\hatbf{n}) F^A_2(\hatbf{n})
		\mathrm{e}^{2\pi i f \Delta\mathbf{x}\cdot\hatbf{n}/c} \dd\hatbf{n}.
	\end{equation}
Here, $F^A_I(\hatbf{n})$ is the antenna response of detector $I$ to gravitational waves of polarization $A$ and $\Delta\mathbf{x}$ is the separation vector between detectors.
The integral is performed over all sky directions $\hatbf{n}$ and a sum is taken over both the ``plus'' and ``cross'' gravitational-wave polarizations.
The leading factor of $5/8\pi$ normalizes the overlap reduction function such that identical, coincident, and co-aligned detectors would have $\gamma(f) = 1$.

Overlap reduction functions are strongly dependent upon baseline geometry -- different pairs of gravitational-wave detectors generically have very different overlap reduction functions.
To illustrate this, the overlap reduction function for the LIGO-Hanford and LIGO-Livingston baseline is shown in blue in Fig. \ref{orfFigure}.
The collection of grey curves, meanwhile, illustrates alternative overlap reduction functions for hypothetical pairs of detectors placed randomly on the surface of the Earth.

The strong dependence of $\gamma(f)$ on baseline geometry raises an interesting possibility.
Given cross-correlation measurements $\hat C(f)$ between two detectors, we can use the measurements themselves to infer the baseline's geometry.
In the electromagnetic domain, a very similar technique has long been used in the field of geodesy: the experimental study of Earth's geometry.
While most commonly used to study the radio sky, very-long baseline interferometry can instead be used to precisely localize radio telescopes on the Earth, allowing for measurements of tectonic motion to better than $\sim 0.1\,\text{mm}\,\text{yr}^{-1}$ \citep{GeodesyA,GeodesyB}.
Similarly, here we will use the \textit{gravitional-wave sky} to determine our detectors' relative positions and orientations.

As an initial demonstration, Fig. \ref{example}a illustrates a simulated observation of the stochastic gravitational-wave background with design-sensitivity Advanced LIGO.
We assume a stochastic energy-density spectrum $\Omega(f) = 3.3\times10^{-9} \left(f/25\,\mathrm{Hz}\right)^{2/3}$, chosen to yield a signal-to-noise ratio (SNR) of 10 after three years of observation.
The dashed curve indicates the mean correlation spectrum $\langle \hat C(f) \rangle$ corresponding to this injection, while the solid trace shows a simulated cross-correlation spectrum $\hat C(f)$ after three years of observation.
By fitting to $\hat C(f)$ (as will be described below in Sect.~\ref{modelSelection}), we can attempt to estimate the geometry of the LIGO Hanford-Livingston baseline.
The resulting posterior on the separation between the LIGO Hanford and Livingston detectors is shown in Fig. \ref{example}b.
This posterior is consistent with the true separation between detectors ($\approx 3000$ km).

\section{Model selection}
\label{modelSelection}

Of course, the physical separations between current gravitational-wave detectors are already known to far better accuracy than can be obtained through gravitational-wave geodesy.
Nevertheless, the ability to measure baseline geometry with the gravitational-wave sky suggests a powerful consistency test for any possible detection of the gravitational-wave background.

In the presence of an isotropic, astrophysical stochastic background, the measured cross-correlation spectrum $\hat C(f)$ \textit{must} exhibit amplitude modulations and zero-crossings consistent with the baseline's overlap reduction function.
Thus, when using the data $\hat C(f)$ to infer the baseline's geometry, we must obtain results that are consistent with the known separations and orientations of the detectors.
In contrast, spurious sources of terrestrial correlation are \textit{not} bound to trace the overlap reduction function.
Hence, there is no \textit{a priori} reason that a correlated terrestrial signal should prefer the true baseline geometry over any other possible detector configuration.

We can more rigorously define this consistency check within the framework of Bayesian hypothesis testing.
Given a measured cross-correlation spectrum $\hat C(f)$, we will ask which of the following hypotheses better describes the data:
\begin{itemize}
\item Hypothesis $\mathcal{H}_\gamma$: The measured cross-correlation is consistent with the true baseline geometry (and hence the baseline's true overlap reduction function).
\item Hypothesis $\mathcal{H}_\mathrm{Free}$: The cross-correlation spectrum is consistent with a model in which we \textit{do not} impose the baseline's known geometry, instead (unphysically) treating the detectors' positions and orientations as free variables to be determined by the data.
\end{itemize}
An isotropic, astrophysical stochastic signal will be consistent with both $\mathcal{H}_\gamma$ and $\mathcal{H}_\mathrm{Free}$ (assuming that the true baseline geometry is among the possible configurations supported in $\mathcal{H}_\mathrm{Free}$).
As the simpler hypothesis, however, $\mathcal{H}_\gamma$ will be favored by the Bayesian ``Occam's factor'' that penalizes the more complex model.
So a true isotropic astrophysical stochastic background will favor $\mathcal{H}_\gamma$.
A generic terrestrial signal, on the other hand, is unlikely to follow the baseline's true overlap reduction function, and so will be better fit by the additional degrees of freedom allowed in $\mathcal{H}_\mathrm{Free}$.
Terrestrial sources of correlation are therefore likely to favor $\mathcal{H}_\mathrm{Free}$.

This procedure is similar to the ``sky scramble'' technique used in pulsar timing searches for very low-frequency gravitational waves \citep{Cornish2016,Taylor2017}.
In pulsar timing experiments, the analogue to the overlap reduction function is the Hellings and Downs curve, which quantifies the expected correlations between pulsars as a function of their angular separation on the sky \citep{HellingsDowns}.
By artificially shifting pulsar positions on the sky, one can seek to disrupt this spatial correlation and produce null data devoid of gravitational-wave signal but that retains other (possibly correlated) noise features.


Given a tentative detection of the stochastic background, we can compute a Bayes factor $\mathcal{B}$ between hypotheses $\mathcal{H}_\gamma$ and $\mathcal{H}_\mathrm{Free}$ to determine which is favored by the data.
Due to the large number of time segments analyzed in stochastic searches, cross-correlation measurements are well-described by Gaussian statistics.
We therefore assume Gaussian likelihoods, such that the probability of measuring $\hat C(f)$ given a model spectrum $C_\mathcal{H}(\Theta;f)$ with parameters $\Theta$ is
	\begin{equation}
	p(\{\hat C\} | \Theta,\mathcal{H}) \propto \exp \left[ 
		- \frac{1}{2} \left(\hat C - C_\mathcal{H}(\Theta) | \hat C - C_\mathcal{H}(\Theta)\right) \right],
	\end{equation}
using the inner product defined in Eq. \eqref{innerProduct}.
	
\begin{figure}
\centering
\includegraphics[width=0.4\textwidth]{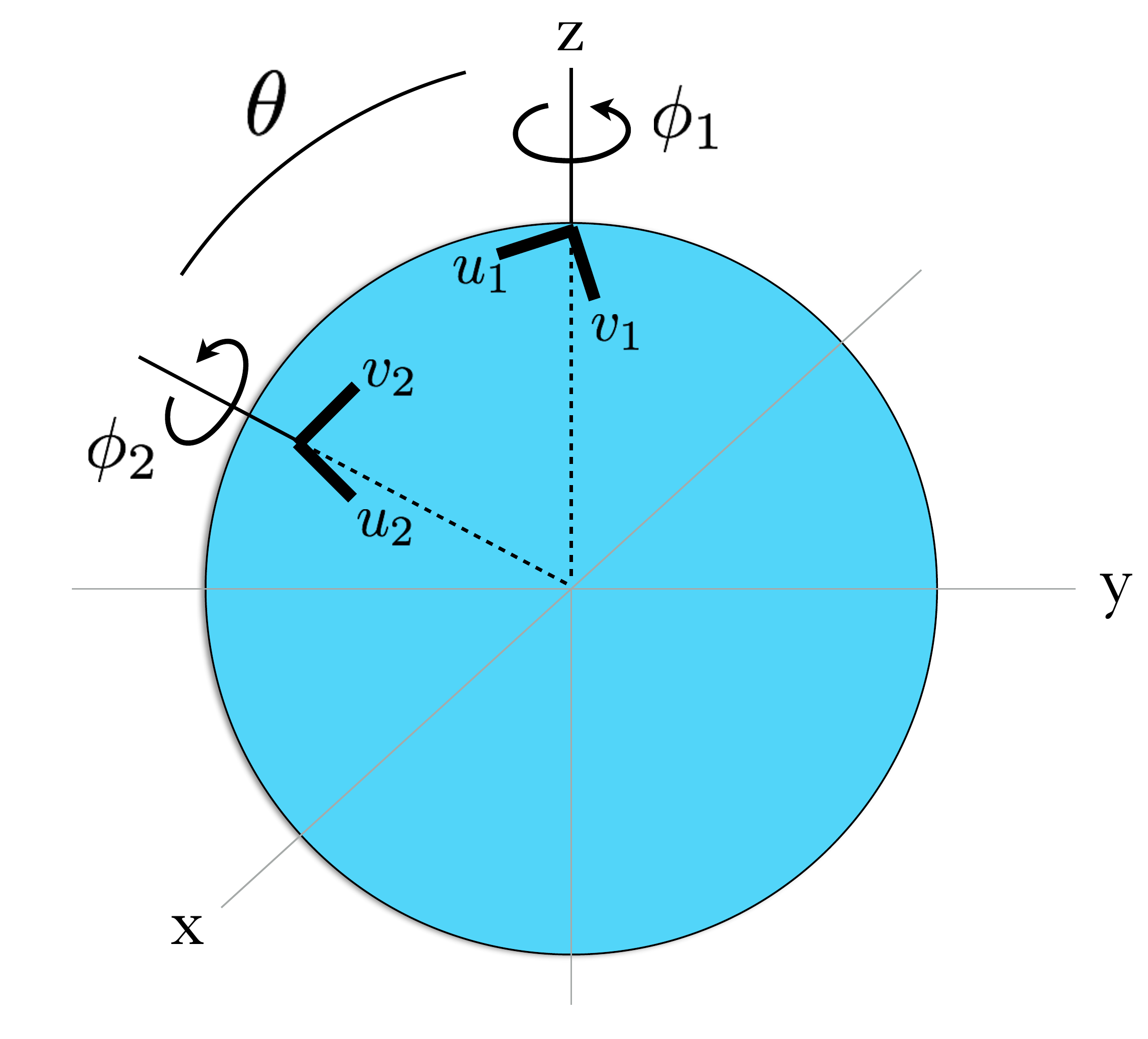}
\caption{
Parametrized geometry of an arbitrary detector baseline on the Earth's surface.
We initially choose coordinates such that the detectors lie in the $x-z$ plane, with one detector at the pole.
The remaining degrees of freedom are the polar angle $\theta$ between detectors, and the rotation angles $\phi_1$ and $\phi_2$ specifying the orientation of each detector.
}
\label{geometry}
\end{figure}

For both hypotheses, we adopt a power-law form for the background's energy-density spectrum, defined by a reference amplitude $\Omega_0$ and a spectral index $\alpha$:
	\begin{equation}
	\label{powerLaw}
	\Omega(f) = \Omega_0 \left(\frac{f}{25\,\text{Hz}}\right)^\alpha.
	\end{equation}
Our model for the cross-correlation spectrum under $\mathcal{H}_\gamma$ is therefore
	\begin{equation}
	\label{GammaC}
	C_\gamma(\Omega_0,\alpha;f) = \gamma_\mathrm{True}(f) \, \Omega_0 \left(f/25\,\text{Hz}\right)^\alpha,
	\end{equation}
where $\gamma_\mathrm{True}(f)$ is the true overlap reduction function for the given baseline.

For $\mathcal{H}_\mathrm{Free}$, we additionally need a parametrized model for possible baseline geometries.	
We use the scheme illustrated in Fig. \ref{geometry}.
Given two detectors on the surface of the Earth (which we approximate as a sphere of radius $R_\oplus = 6.4\times 10^6$ m), one can choose coordinates such that the first detector lies at the pole and the second along the meridian (in the $x-z$ plane).
We then have three remaining degrees of freedom:
the polar angle $\theta$ between detectors and the angles $\phi_1$ and $\phi_2$ specifying the rotation of each detector about its local zenith.
Specifically, $\phi_{1/2}$ are the angles between the detectors' $\hat v$ arms (see Fig. \ref{geometry}) and the $y$-axis.
For convenience, below we will work in terms of the distance $\Delta x = 2 R_\oplus \sin \theta/2$ between detectors, rather than the polar angle.
All together, the model cross-correlation spectrum under hypothesis $\mathcal{H}_\mathrm{Free}$ is
	\begin{equation}
	\label{FreeC}
	C_\text{Free}(\Omega_0,\alpha,\Delta x,\phi_1,\phi_2; f) = \gamma(\Delta x,\phi_1,\phi_2;f) \, \Omega_0 \left(f/25\,\text{Hz}\right)^\alpha.
	\end{equation}
	

We choose a log-uniform prior on $\Omega_0$ between $(10^{-12},10^{-6})$ (extending well above and well below Advanced LIGO's sensitivity) and uniform priors on $\phi_1$ and $\phi_2$ on $(0,2\pi)$.
Similarly, we use a uniform prior on $\cos\theta$ between $(-1,1)$, corresponding to a prior $p(\Delta x) \propto \Delta x$ on the distance between detectors.
We adopt a triangular prior on the background's spectral index: $p(\alpha)\propto 1-|\alpha|/\alpha_\text{Max}$, with $\alpha_\text{Max}=6$.
This prior penalizes very steeply-sloped backgrounds, while still accommodating backgrounds much steeper than those predicted from known sources.

\section{Demonstration}
\label{demonstration}

To explore our ability to differentiate terrestrial correlation from an astrophysical background, we will simulate Advanced LIGO measurements of three different sources of correlation:
an isotropic stochastic background, a correlated frequency comb, and magnetic Schumann resonances.
These latter two sources are terrestrial, and hence should be disfavor $\mathcal{H}_\gamma$ over $\mathcal{H}_\mathrm{Free}$.
We note that there exist dedicated strategies for identifying and mitigating combs and Schumann resonances \citep{Covas2018,Thrane2014}.
Here, we use combs and Schumann resonances simply as proxies for any as-of-yet \textit{unknown} sources of terrestrial correlation that could contaminate stochastic search efforts.

\begin{figure}
\centering
\includegraphics[width=0.48\textwidth]{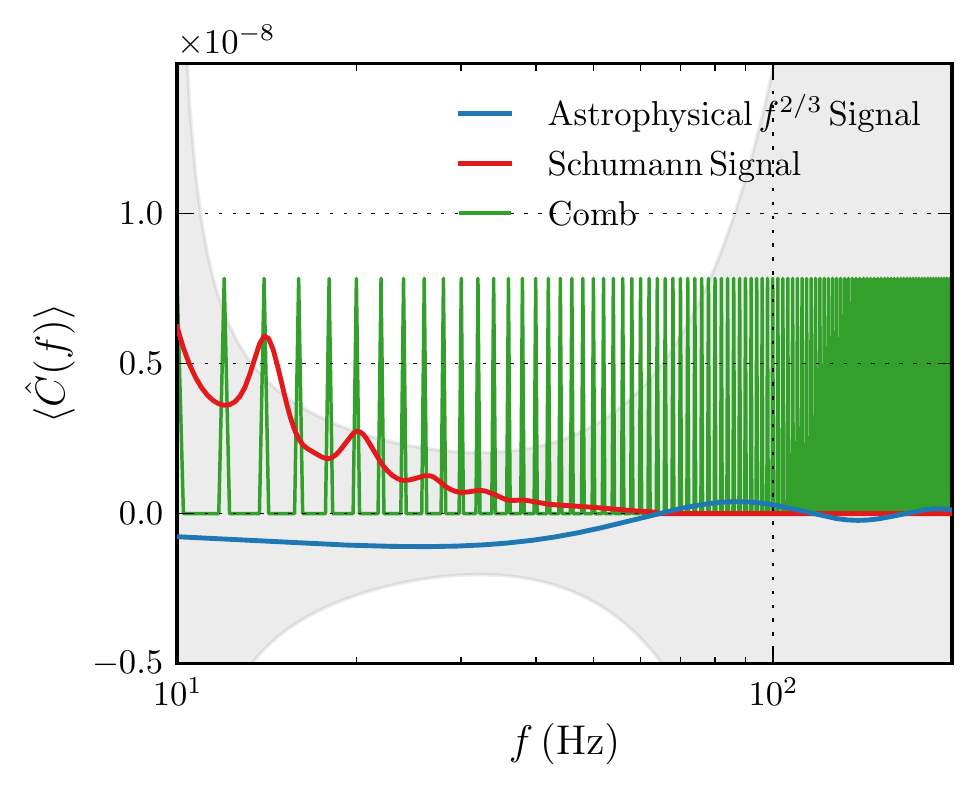}
\caption{
Mean cross-correlation spectra used to simulate stochastic search measurements with the Advanced LIGO Hanford and Livingston detectors.
We consider an isotropic astrophysical stochastic background, with energy density $\Omega(f)\propto f^{2/3}$ [blue; Eq. \eqref{isotropicModel}].
We additionally consider two sources of terrestrial, non-astrophysical correlation: a signal due to magnetic Schumann resonances [red; Eq. \eqref{schumannModel}] and a correlated frequency comb with $\Delta f = 2$ Hz spacing [green; Eq. \eqref{combModel}].
The amplitudes of the spectra have been scaled such that each is expected to be detected with a signal-to-noise ratio of 10 after three years of observation with design-sensitivity Advanced LIGO.
For comparison, the grey band illustrates the $\pm1\sigma$ uncertainties of a cross-correlation search after three years of integration.
}
\label{injection}
\end{figure}

\begin{figure}
\centering
\includegraphics[width=0.48\textwidth]{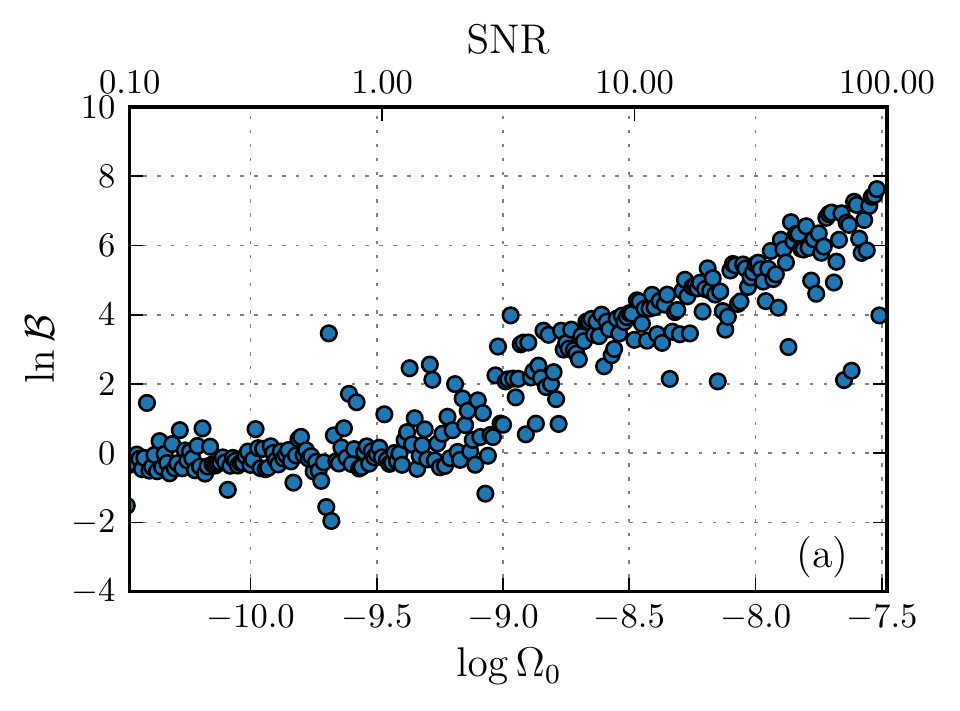}
\includegraphics[width=0.48\textwidth]{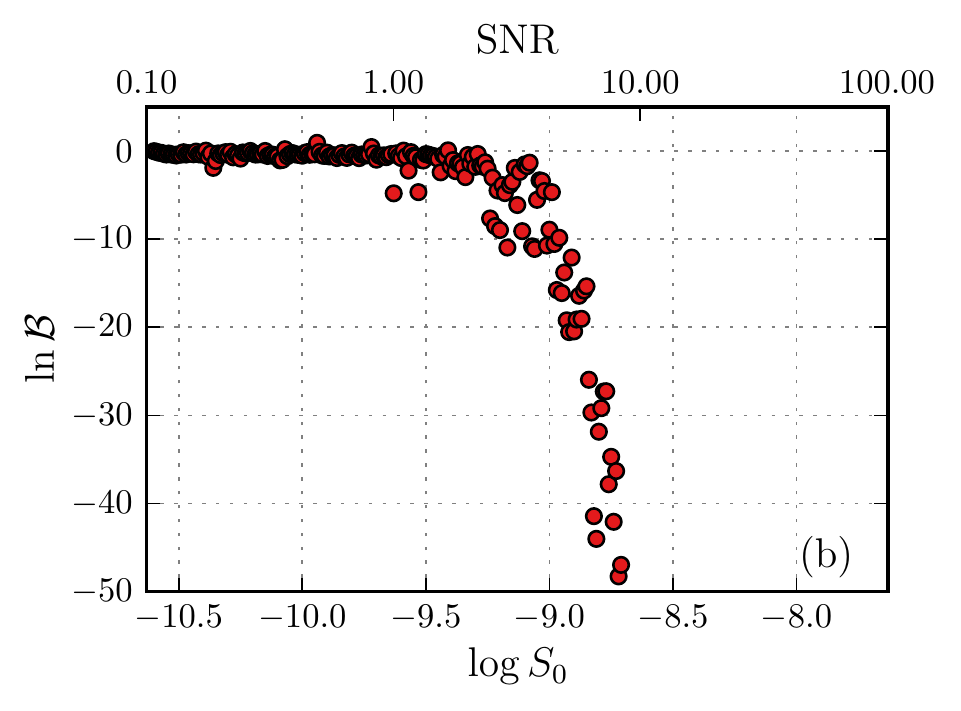}
\includegraphics[width=0.48\textwidth]{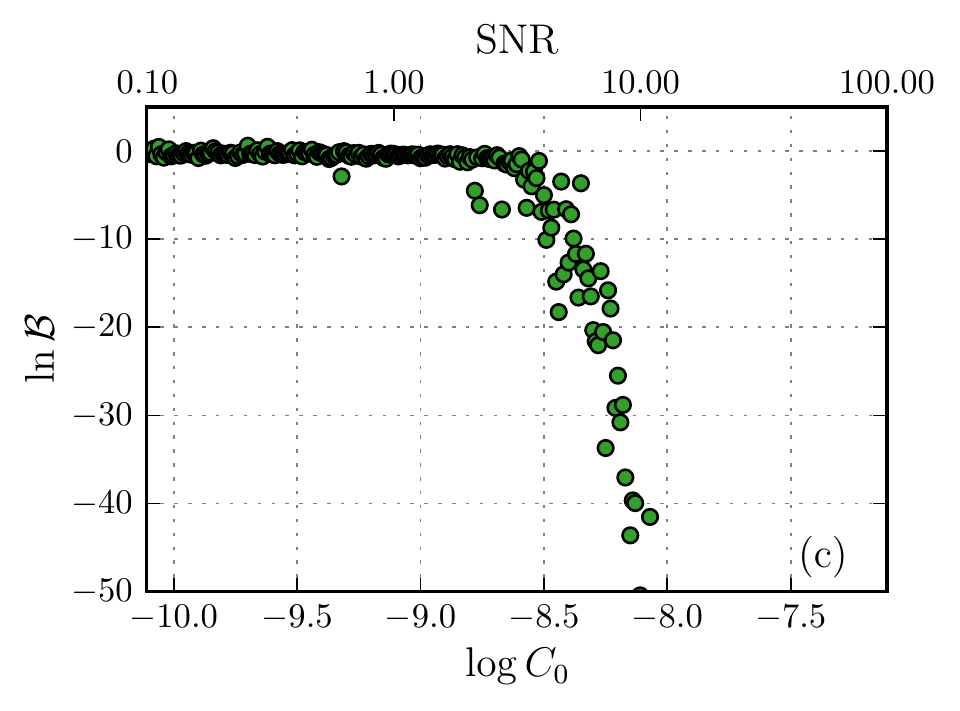}
\caption{
Log-Bayes factors between the physical and unphysical hypotheses $\mathcal{H}_\gamma$ and $\mathcal{H}_\mathrm{Free}$ as a function of injection strength for isotropic astrophysical backgrounds, Schumann resonances, and correlated combs [Eqs. \eqref{isotropicModel}, \eqref{combModel}, and \eqref{schumannModel}].
To enable a direct comparison between injection types, the upper horizontal axes show the signal-to-noise ratios of these injections.
$\ln\mathcal{B}$ increases linearly with the strength of an astrophysical injection, indicating consistency with the correct (known) detector geometry.
Meanwhile, $\ln\mathcal{B}$ decreases exponentially for the terrestrial sources of correlation, \textit{disfavoring} the correct geometry.
In these cases, at least, $\ln\mathcal{B}$ therefore successfully discriminates between astrophysical and terrestrial sources of measured cross-correlation.
}
\label{bayes}
\end{figure}

Below, we describe the model cross-correlation spectra adopted for each test case:

\textit{1. Isotropic stochastic gravitational-wave background:}
We assume that the stochastic gravitational-wave background is well-described by a power-law with spectral index $\alpha=2/3$, as predicted for compact binary mergers.
The corresponding expected cross-correlation spectrum is
	\begin{equation}
	\langle C(f)\rangle_\text{Stoch} = \gamma_\text{LIGO}(f) \,\Omega_0 \left(\frac{f}{25\,\text{Hz}}\right)^{2/3},
	\label{isotropicModel}
	\end{equation}
where $\gamma_\text{LIGO}(f)$ is the overlap reduction function for the Hanford-Livingston baseline (shown in Fig. \ref{orfFigure}).

\textit{2. Frequency comb:}
We consider a correlated comb of uniformly-spaced lines, separated in frequency by $\Delta f$ and with heights set by $C_0$:
	\begin{equation}
	\langle C(f) \rangle_\text{Comb} = C_0 \Delta f \sum_{n=0}^{\infty} \delta (f - n \Delta f).
	\label{combModel}
	\end{equation}
Note that the leading factor of $\Delta f$ in Eq. \eqref{combModel} ensures that $C_0$ is dimensionless.
In the examples below, we use a comb spacing of $\Delta f = 2$ Hz.

\textit{3. Magnetic Schumann resonances:}
Given an environmental magnetic field $\tilde m(f)$, the strain induced in a gravitational-wave detector is $\tilde s(f) = T(f) \tilde m(f)$, where $T(f)$ is a transfer function with units $[\text{strain}/\text{Tesla}]$.
If there exists a correlated magnetic power spectrum $M(f) = \langle \tilde m^*_1(f) \tilde m_2(f) \rangle$ between the sites of two gravitational-wave detectors, then from Eq. \eqref{crosscorr} the resulting strain correlation will be of the form $\hat C(f) \propto f^3 |T(f)|^2 \,\mathrm{Re}\,M(f)$.
We take $M(f)$ to be the median Schumann auto-power spectrum measured at the Hylaty station in Poland, as reported in Ref. \citep{Coughlin2018}.
This may not exactly match the magnetic cross-power spectrum between LIGO-Hanford and LIGO-Livingston.
Most notably, we take $\mathrm{Re}\,M(f)$ to be everywhere positive, as the (potentially frequency-dependent) sign of the Schumann cross-power between the LIGO detectors is not well-known.
Nevertheless, this model captures the qualitative features expected of a Schumann signal.
The magnetic transfer functions for the LIGO detectors are expected to be power-laws, but their spectral indices are also not well-known; we somewhat arbitrarily choose $T(f) \propto f^{-2}$.
Our Schumann signal model is therefore
	\begin{equation}
	\langle C(f) \rangle_\text{Schumann} = S_0 \left(\frac{f}{25\,\mathrm{Hz}}\right)^{-1} \frac{\mathrm{Re}\,M(f)}{\mathrm{Re}\,M(25\,\mathrm{Hz})},
	\label{schumannModel}
	\end{equation}
normalized so that $S_0$ is the cross-correlation measured at the reference frequency $25$\,Hz.

The mean cross-correlation spectra for the astrophysical, Schumann, and comb models are shown in Fig. \ref{injection}.
For each source of correlation, we simulate Advanced LIGO measurements of three hundred injected signals, with expected signal-to-noise ratios ranging from 0.1 to 100.
To produce each realization, we scale the amplitude parameters ($\Omega_0$, $C_0$, and $S_0$) to obtain the desired SNR and add random Gaussian measurement noise $\delta C(f)$ with variance given by Eq. \eqref{sigma}.
For each simulated measurement, we compute a Bayes factor $\mathcal{B}$ between $\mathcal{H}_\gamma$ and $\mathcal{H}_\mathrm{Free}$ to determine whether the data physically favors the correct detector geometry, or unphysically favors some alternate geometry.
We compute Bayesian evidences using \texttt{MultiNest} \citep{Feroz2008,Feroz2009}, an implementation of the nested sampling algorithm \citep{Skilling2004,Skilling2006}.
We make use of \texttt{PyMultiNest}, which provides a Python interface to \texttt{MultiNest} \citep{Buchner2014}.

The resulting Bayes factors are plotted in Fig. \ref{bayes} as a function of injected signal amplitude.
As physically distinct parameters, the power-law, Schumann, and comb amplitudes should not be directly compared to one another.
Instead, we show the injections' expected signal-to-noise ratios (which \textit{can} be directly compared) on the upper horizontal axes.
To compute these SNRs, we assume recovery with a power-law model of slope $\alpha=2/3$.
Thus the SNRs of the power-law injections are optimal.
While SNRs for the comb and Schumann injections are \textit{not} optimal (as the recovery model and injections are not identical), they do represent the signal-to-noise ratios at which such signals would contaminate searches for the stochastic background.

At signal-to-noise ratios $\mathrm{SNR}\lesssim1$, the log-Bayes factors for all three sources of correlation cluster near $\ln\mathcal{B}\sim 0$.
For an astrophysical signal above $\mathrm{SNR}\sim1$, $\ln\mathcal{B}$ becomes positive, growing approximately linearly with $\log\Omega_0$.
In contrast, $\ln\mathcal{B}$ falls exponentially to large negative values as we increase the amplitude of Schumann and comb injections.
In Appendix \ref{laplace}, we illustrate how the Laplace approximation can be used to derive these approximate scaling relations.

It is instructive to look at parameter estimation results for specific astrophysical, comb, and Schumann injections.
In Appendix \ref{peResults}, we show posteriors on the parameters of $\mathcal{H}_\mathrm{Free}$ obtained using simulated observations of the stochastic background, a frequency comb, and a Schumann signal.
As suggested by Fig. \ref{bayes}, an observation of an isotropic stochastic background yields posteriors consistent with Advanced LIGO's correct geometry.
The comb and Schumann observations, on the other hand, produce unphysical posteriors on the positions and orientations of the Advanced LIGO detectors.


Figure \ref{bayes} demonstrates that gravitational-wave geodesy can be used to successfully reject cross-correlation spectra that are inconsistent with Advanced LIGO's overlap reduction function.
There nevertheless remains the possibility of false positives: non-astrophysical correlation spectra that, purely by chance, yield posteriors consistent with Advanced LIGO's geometry.
To carefully calculate the probability of a false positive at a particular $\mathcal{B}$, one could analyze a set of random cross-correlation spectra (e.g. drawn from the space of spectra supported by $\mathcal{H}_\mathrm{Free}$) and construct a null distribution of the resulting Bayes factors.
Alternatively, we can quickly estimate the probability of false positives at a given $\ln\mathcal{B}$ using Fig. \ref{bayes}.a.
Given equal prior odds for $\mathcal{H}_\gamma$ and $\mathcal{H}_\mathrm{Free}$, the Bayes factors in Fig. \ref{bayes}\,a may be directly interpreted as odds ratios.
A Bayes factor of $\ln\mathcal{B}=4$ (corresponding to $\mathrm{SNR}\approx10$), for example, indicates $e^4:1$ odds that the given data is drawn from $\mathcal{H}_\gamma$ versus $\mathcal{H}_\mathrm{Free}$.
If taken at face value, this implies that if we were to simulate $e^4+1\approx 56$ random correlation spectra compatible with $\mathcal{H}_\mathrm{Free}$ and with $\mathrm{SNR}=10$, then we should expect \textit{one} of the spectra to yield $\ln\mathcal{B}\gtrsim 4$ by chance.
In this way, our formalism not only offers a means of rejecting non-astrophysical correlations, but can bolster the statistical significance of a real stochastic signal.

	
\section{Discussion and Conclusion}

As searches for the stochastic gravitational-wave background grow increasingly sensitive, we may be nearing the first detection of the background.
This prospect, though, comes with significant risk, namely the high cost of a false-positive detection.
To minimize this risk, it will be important to develop methods to validate tentative detections of the gravitational-wave background.
Specifically, when assessing any apparent detection, it will be necessary to argue not just that an observed correlation is statistically-significant, but that it it \textit{astrophysical} -- that it is due to gravitational waves and not some other, terrestrial process.
While well-developed methods exist to quantify the statistical significance of measured correlations, until now no generic method has existed to gauge whether a statistically significant cross-correlation is indeed astrophysical.

In this paper, we explored how gravitational-wave geodesy -- the use of the stochastic gravitational-wave background itself to determine the positions and orientations of gravitational-wave detectors -- can form the basis for a novel consistency check on an apparent detection of the background.
If the measured correlation between detectors truly represents a gravitational-wave signal, then the reconstructed detector orientations and positions must be compatible with their true, known values.
Correlations due to any terrestrial source, on the other hand, have no reason to prefer the baseline's true geometry over any other possible arrangement.
By demanding that gravitational-wave geodesy yield results consistent with the true baseline geometry, we can discriminate between astrophysical and terrestrial sources of correlation.
Used in this fashion, gravitational-wave geodesy provides a second independent measure of detection significance besides a standard signal-to-noise ratio.

Our analysis has relied on two important assumptions.
First, we have adopted a relatively simple model energy-density spectrum (a power law) that was a good description of our simulated stochastic signals.
In Appendix \ref{complicationsA}, we investigate how our method performs given more complex models for the stochastic background.
Most importantly, we also investigate how are results are affected if we mistakenly assume an \textit{incorrect} form for the background's energy-density spectrum.
We find that remains robust, correctly classifying astrophysical signals even given significant mismatch between our model spectrum and the true stochastic signal.

Second, we have assumed that the stochastic gravitational-wave background is isotropic, which is unlikely to be strictly true.
As discussed further in Appendix \ref{complicationsB}, the expected anisotropies in the stochastic background are small, and therefore are unlikely to affect our analysis.
In the case that anisotropy is a significant concern, however, we outline how our analysis should be modified to handle an anisotropic stochastic signal.

\acknowledgements{\textbf{Acknowledgements}}

We would like to thank Sharan Banagiri, Jan Harms, Andrew Matas, Joe Romano, Colm Talbot, Steve Taylor, Eric Thrane, Alan Weinstein, and members of the LIGO/Virgo Stochastic Data Analysis Group for useful comments and conversation.
LIGO was constructed by the California Institute of Technology and Massachusetts Institute of Technology with funding from the National Science Foundation and operates under cooperative agreement PHY-0757058 .
MC was supported by the David and Ellen Lee Postdoctoral Fellowship at the California Institute of Technology.
This paper carries LIGO Document Number LIGO-P1800226.

\bibliographystyle{aasjournal}
\bibliography{references}

\appendix
\section{Bayes Factor Scaling}
\label{laplace}

The behavior of the Bayes factors in Fig. \ref{bayes} can be understood using the Laplace approximation.
The Laplace approximation involves the following two assumptions:
	\begin{itemize}
	\item Our prior $p(\Theta|\mathcal{H})$ on the parameters of hypothesis $\mathcal{H}$ is flat over a range $\Delta \Theta$, so that $p(\Theta|\mathcal{H}) = 1/\Delta\Theta$.
	\item The likelihood $p(\hat C | \Theta, \mathcal{H})$ is strongly peaked about maximum-likelihood parameter values $\overbar\Theta$ and a peak value $\overbar{\mathcal{L}}$. The width of the peak is $\delta\Theta$.
	\end{itemize}
Under these assumptions, a Bayesian evidence may be approximated as
	\begin{equation}
	\begin{aligned}
	p(\hat C | \mathcal{H}) &= \int p(\hat C | \Theta,\mathcal{H}) p(\Theta | \mathcal{H}) d\Theta \\
		&\approx \frac{\delta \Theta}{\Delta\Theta} \overbar{\mathcal{L}}.
	\end{aligned}
	\end{equation}
The leading term $\delta\Theta/\Delta\Theta$ can be interpreted as the volume of the available parameter space that is compatible with the measured data.
Given two hypotheses $\mathcal{H}_A$ and $\mathcal{H}_B$, the Bayes factor between them becomes
	\begin{equation}
	\begin{aligned}
	\mathcal{B}^A_B 
		&= \frac{ p(\hat C|\mathcal{H}_A) } { p(\hat C |\mathcal{H}_B)} \\
		& \approx \frac{\delta\Theta_A/\Delta\Theta_A}{\delta\Theta_B/\Delta\Theta_B}
			\frac{\overbar{\mathcal{L}}_A}{\overbar{\mathcal{L}}_B}.
	\end{aligned}
	\end{equation}
The ratio $\overbar{\mathcal{L}}_A/\overbar{\mathcal{L}}_B$ is the standard maximum likelihood ratio between $\mathcal{H}_A$ and $\mathcal{H}_B$.
The preceding term, known as the ``Occam's factor,'' penalizes the more complex hypothesis with the larger available parameter space.
Using the Laplace approximation, our Bayes factor between hypotheses $\mathcal{H}_\gamma$ and $\mathcal{H}_\mathrm{Free}$ may be written
	\begin{equation}
	\label{Laplace}
	\begin{aligned}
	\mathcal{B} &= \frac{p(\hat C\ | \mathcal{H}_\gamma)}{p(\hat C | \mathcal{H}_\mathrm{Free})} \\
		&\approx \left[\frac{\delta\Omega_0}{\Delta\Omega_0}
				\frac{\delta\alpha}{\Delta\alpha}\right]_\gamma
			\left[\frac{\delta\Omega_0}{\Delta\Omega_0}
				\frac{\delta\alpha}{\Delta\alpha}
				\frac{\delta\theta}{\Delta\theta}
				\frac{\delta\phi_1}{\Delta\phi_1}
				\frac{\delta\phi_2}{\Delta\phi_2}\right]^{-1}_\mathrm{Free} \\
			&\hspace{1cm}\times
				\frac{\exp\left[-\frac{1}{2}\left(\hat C - \overbar C_\gamma |\hat C - \overbar C_\gamma\right) \right]}
					{\exp\left[-\frac{1}{2}\left(\hat C - \overbar C_\mathrm{Free} |\hat C - \overbar C_\mathrm{Free}\right) \right]},
	\end{aligned}
	\end{equation}
where $\overbar C_\gamma$, for instance, is the maximum-likelihood fit to the data under the $\mathcal{H}_\gamma$ hypothesis.

First, consider the case of an isotropic astrophysical background of amplitude $\Omega_0$.
In this case, both hypotheses $\mathcal{H}_\gamma$ and $\mathcal{H}_\mathrm{Free}$ can successfully fit the resulting cross-correlation spectrum.
Then $\hat C - \overbar C_\gamma \approx \hat C - \overbar C_\mathrm{Free} \approx 0$ and the likelihood ratio in Eq. \eqref{Laplace} is approximately one.
Because both models can fit the data, posteriors on each parameter (of each hypothesis) are peaked, with fractional widths that scale (e.g. $\delta\theta/\Delta\theta$) that scale as $\mathrm{SNR}^{-1}\propto\Omega_0^{-1}$.
Then, in the presence of an astrophysical background, we expect Eq. \eqref{Laplace} to scale as $\mathcal{B}\propto  \Omega_0^3$, or
	\begin{equation}
	\ln\mathcal{B}\sim 3\log\Omega_0,
	\end{equation}
up to additive constants.

Next, consider a correlated signal of terrestrial origin, characterized by some amplitude $C_0$.
We will assume that $\mathcal{H}_\gamma$ is unable to accommodate the measured correlations, but that $\mathcal{H}_\mathrm{Free}$, with a greater number of free parameters, \textit{can} successfully fit the data to some extent.
Then $\hat C - \overbar C_\mathrm{Free}\approx 0$ but $\hat C-\overbar C_\gamma\ne 0$.
So the likelihood term in Eq. \eqref{Laplace} is not constant, but will depend exponentially on $C_0$.
Ignoring the leading Occam's factors (which can scale at most as a power law in $C_0$), our Bayes factor becomes
	\begin{equation}
	\begin{aligned}
	\mathcal{B} &\propto \exp\left[-\frac{1}{2}\left(\hat C-\overbar C_\gamma | \hat C-\overbar C_\gamma\right)\right] \\
		&\propto \exp\left[-\frac{1}{2}\left(\hat C | \hat C\right) + \left(\hat C|\overbar C_\gamma\right)
			- \frac{1}{2}\left(\overbar C_\gamma | \overbar C_\gamma\right)\right],
	\end{aligned}
	\end{equation}
giving
	\begin{equation}
	\label{bayesTemp}
	\ln\mathcal{B} \sim -\frac{1}{2}\left(\hat C | \hat C\right) + \left(\hat C|\overbar C_\gamma\right)
			- \frac{1}{2}\left(\overbar C_\gamma | \overbar C_\gamma\right).
	\end{equation}
The maximum likelihood value of $\Omega_0$ [the amplitude of our model spectrum $C_\gamma(f)$] is given by \citep{Callister2016}
	\begin{equation}
	\overbar  \Omega_0 = \frac{\left(f^{2/3} | \hat C\right)}{\left(f^{2/3} | f^{2/3} \right)}.
	\end{equation}
Although this does scale proportionally with $C_0$, in this scenario our measured correlation $\hat C(f)$ is assumed to have a very different shape from an astrophysical power law.
The inner product $(f^{2/3} | \hat C)$ may therefore be small, in which case the cross term $(\hat C|\overbar C_\gamma)$ in Eq. \eqref{bayesTemp} may be neglected.
As a result, $\ln\mathcal{B} \propto - C_0^2$, or
	\begin{equation}
	\ln\mathcal{B} \propto - 10^{2\log C_0}.
	\end{equation}
	
\section{Parameter Estimation Results}
\label{peResults}

\begin{figure*}[ht]
\centering
\includegraphics[width=0.47\textwidth]{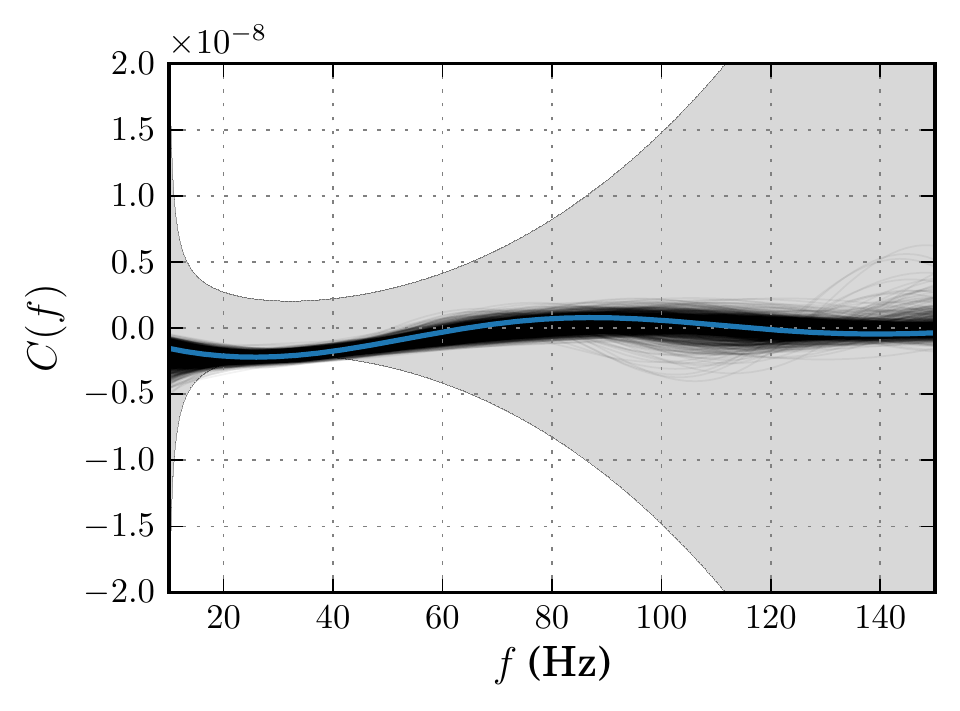} \\
\includegraphics[width=0.47\textwidth]{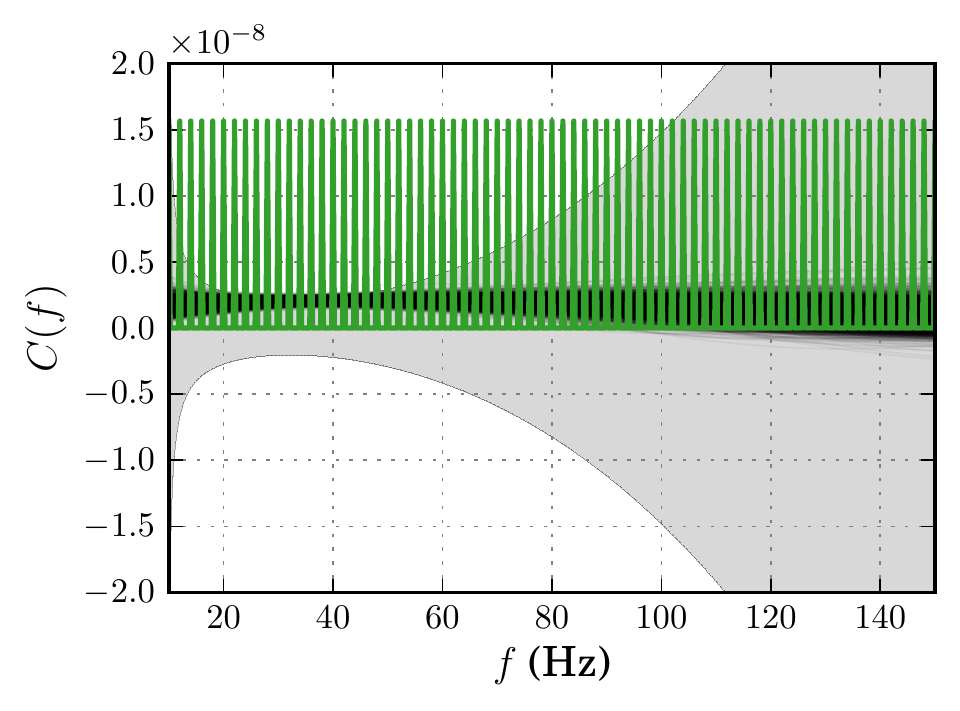} \\
\includegraphics[width=0.47\textwidth]{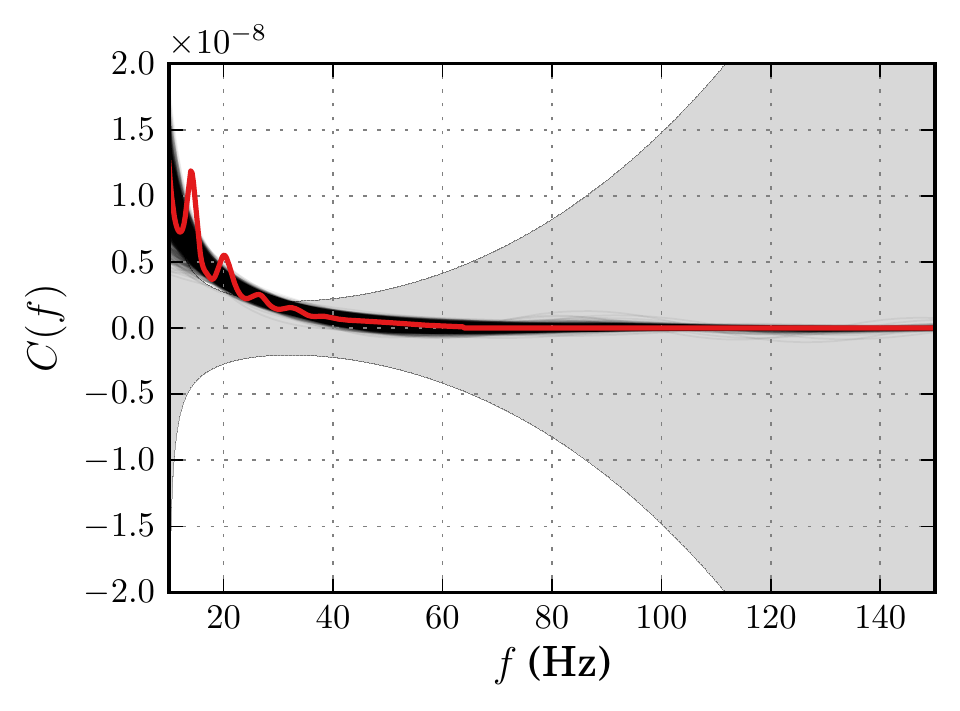}
\caption{Reconstructed cross-correlation spectra using simulated Advanced LIGO observations of an isotropic gravitational-wave background (\textit{top}), a correlated frequency comb (\textit{middle}), and Schumann resonances (\textit{bottom}).
The blue, green, and red curves show the injected gravitational-wave, comb, and Schumann spectra, respectively, while the shaded bands indicate the $\pm 1\sigma$ uncertainty region on the simulated measurements.
We perform parameter estimation on each injection using the $\mathcal{H}_\mathrm{Free}$ hypothesis, fitting simultaneously for the spectral shape of a presumed stochastic background as well as the detectors' separation and orientations.
The collections of grey curves show the resulting posteriors on the injected cross-correlation spectrum.
Posteriors on the model parameters themselves are shown in Figs. \ref{examplePowerLaw}-\ref{exampleSchumann}.
}
\label{examplePowerLawReconstruction}
\end{figure*}

\begin{figure*}
\centering
\includegraphics[width=0.95\textwidth]{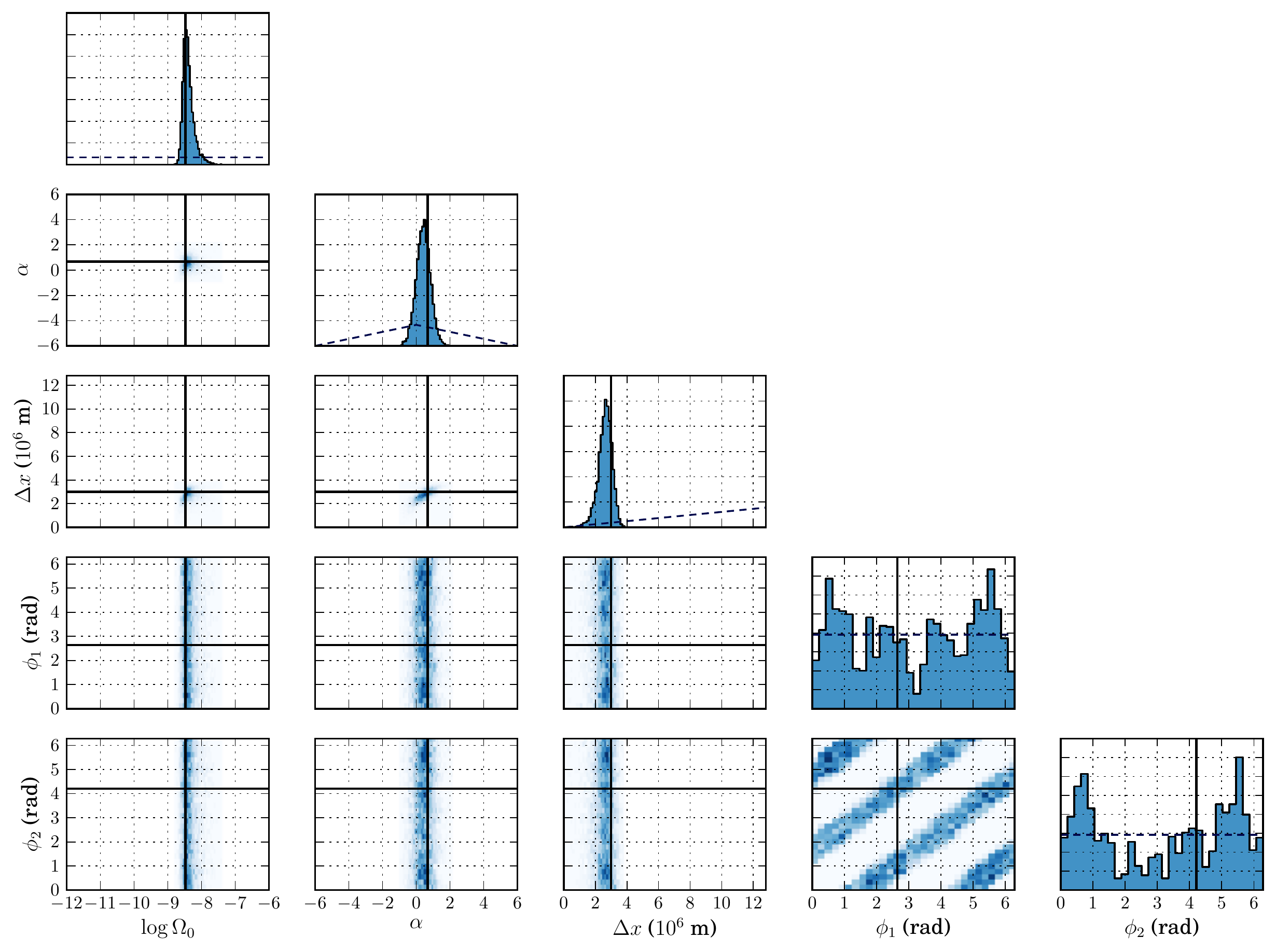}
\caption{Example posterior on the stochastic background amplitude $\Omega_0$ and spectral index $\alpha$, as well the separation $\Delta x$ and rotation angles $\phi_1$ and $\phi_2$ of the Advanced LIGO detectors, given a simulated three-year observation of an isotropic astrophysical background.
The injected signal has spectral index $\alpha=2/3$ and amplitude $\Omega_0 = 3.33\times10^{-9}$, with an expected signal-to-noise ratio of 10.
Dashed lines in the one-dimensional marginalized posteriors show the prior adopted for each parameter, while solid black lines mark the injected background parameters and the true Advanced LIGO geometry.
In addition to recovering the amplitude and spectral index of the injected stochastic signal, we obtain posteriors consistent with the true separation and rotation angles of the Advanced LIGO detectors.
}
\label{examplePowerLaw}
\end{figure*}

\begin{figure*}
\centering
\includegraphics[width=0.95\textwidth]{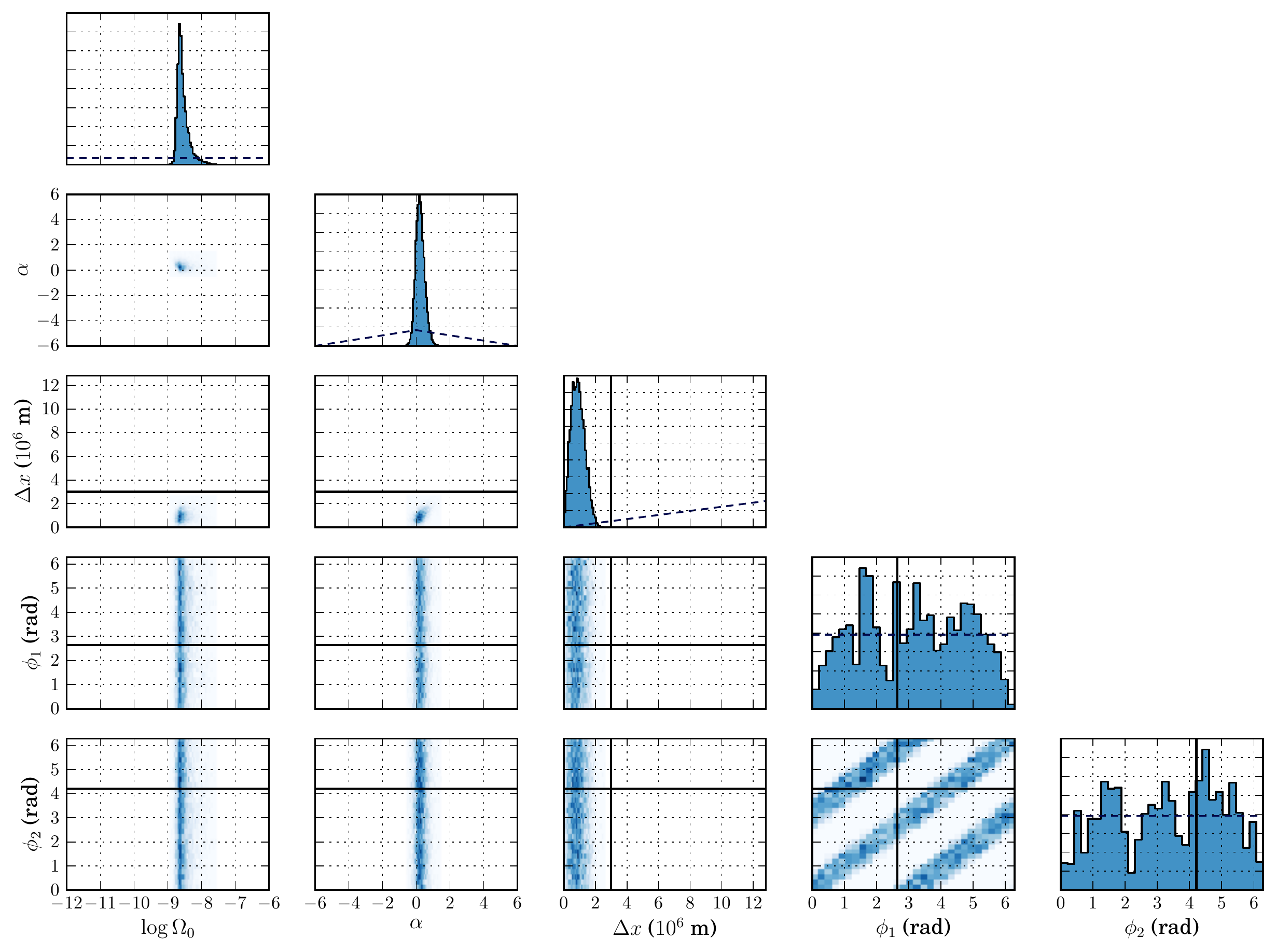}
\caption{As in Fig. \ref{examplePowerLaw} above, but for a simulated measurement of a correlated frequency comb with spacing $\Delta f = 2$ Hz and height $C_0 = 7.83\times10^{-9}$.
The comb's amplitude is chosen so that it has $\mathrm{SNR}=10$ after three years of observation.
The correlated comb is not well fit by the Advanced LIGO overlap reduction function, and so our recovered posteriors on Hanford and Livingston's separation and rotation angles are inconsistent with their known values (solid black lines).
}
\label{exampleComb}
\end{figure*}

\begin{figure*}
\centering
\includegraphics[width=0.95\textwidth]{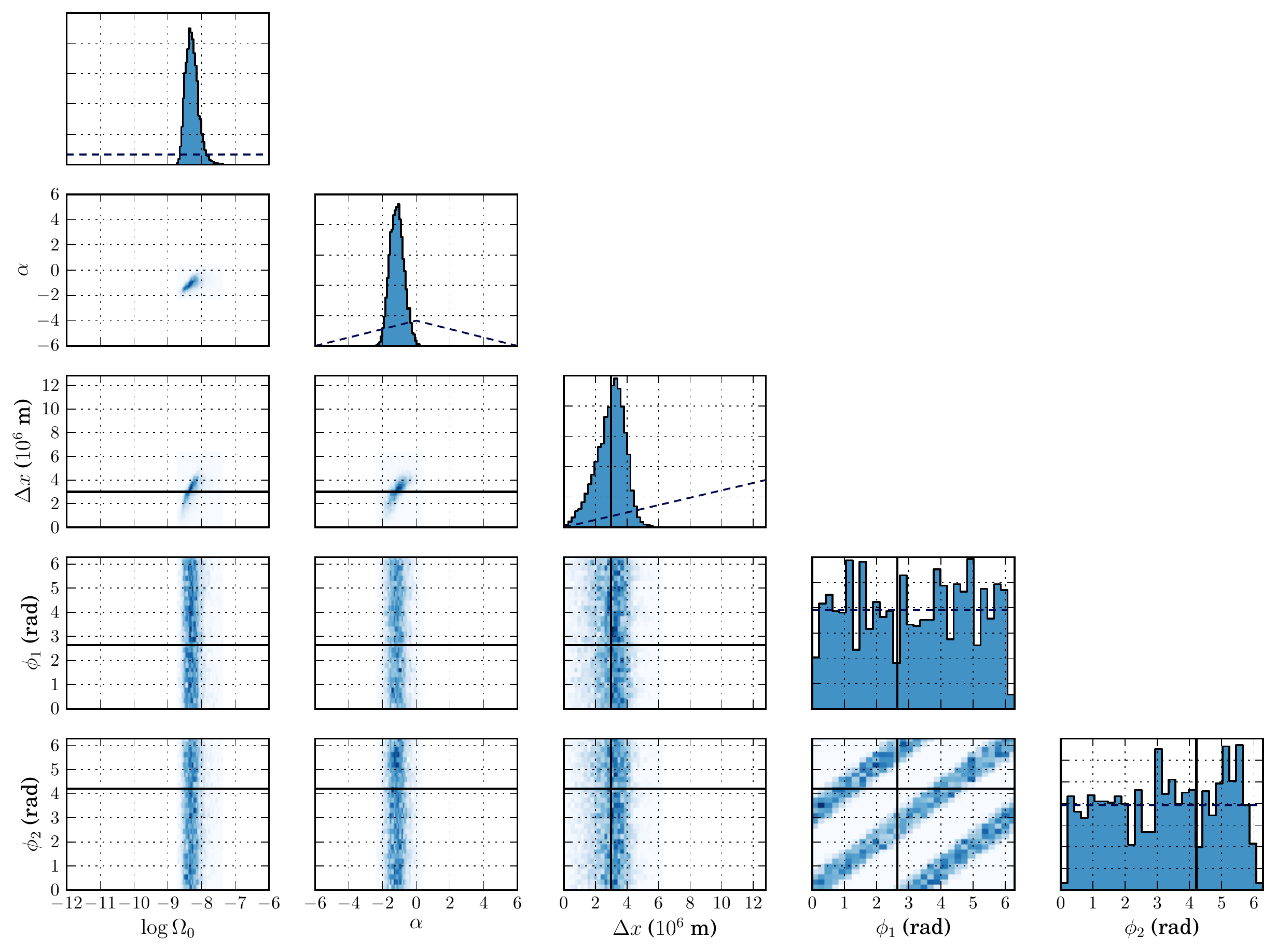}
\caption{As in Figs. \ref{examplePowerLaw} and \ref{exampleComb} above, for a simulated observation of a correlated Schumann signal of height $S_0=2.33\times10^{-9}$, chosen to yield $\mathrm{SNR}=10$ after three years of observation.
While the posterior does encompass the correct Hanford-Livingston separation, it is incompatible with the detectors' true rotation angles.}
\label{exampleSchumann}
\end{figure*}

In this section we show example parameter estimation results obtained when analyzing simulated observations of a stochastic gravitational-wave background, a correlated frequency comb, and Schumann resonances, each with $\mathrm{SNR}=10$.
For each injection, we perform parameter estimation under the $\mathcal{H}_\mathrm{Free}$ hypothesis, allowing the detector positions and orientations to (unphysically) vary to best match the observed cross-correlation spectrum.
We implement parameter estimation using \texttt{MultiNest} and \texttt{PyMultiNest}.

Figure \ref{examplePowerLawReconstruction} shows the three injections as well as the posteriors obtained on each cross-correlation spectrum.
With the five free parameters afforded by $\mathcal{H}_\mathrm{Free}$, we succeed in reasonably fitting each of the three spectra.
Note that, although we appear to poorly recover the correlated comb injection, the posterior on $C(f)$ closely matches the \textit{frequency-averaged} correlation.

Although the gravitational-wave background, comb, and Schumann injections are all reasonably well-fit under $\mathcal{H}_\mathrm{Free}$, they yield very different posteriors on Advanced LIGO's baseline length $\Delta x$ and detector orientation angles $\phi_1$ and $\phi_2$.
Fig. \ref{examplePowerLaw} shows the parameter posteriors given by the simulated gravitational-wave background.
The diagonal subplots show marginalized one-dimensional posteriors on each parameter, while the central subplots show joint posteriors between each pair of parameters.
The solid black lines indicate true parameter values and dashed curves show the priors placed on each parameter.
We recover posteriors consistent with the amplitude and spectral index of the injected stochastic signal.
More importantly, we also obtain well-behaved posteriors on Advanced LIGO's geometry, with a distance posterior (the same as shown in Fig. \ref{example}) consistent with the true separation between detectors.
Interestingly, although neither $\phi_1$ nor $\phi_2$ are well-constrained, their \textit{difference} is well-measured.
This can be seen in the joint posterior between both angles, which strongly supports diagonal bands of constant $\phi_1 - \phi_2$, including the true rotation angles of Hanford and Livingston.
We therefore have strong support for the correct detector geometry, yielding a log-Bayes factor $\ln\mathcal{B} = 3.6$ ($\mathcal{B}=36.6$) in favor of $\mathcal{H}_\gamma$.

Fig. \ref{exampleComb}, meanwhile, shows parameter estimation results obtained for the comb injection.
As seen in Fig. \ref{examplePowerLawReconstruction} above, we have enough freedom to fit the (average) cross-correlation spectrum, yielding reasonably-peaked posteriors in Fig. \ref{exampleComb}.
However, the posteriors on detector separation and orientation are unphysical, excluding the known Hanford-Livingston geometry.
We therefore obtain $\ln\mathcal{B} = -58.5$ ($\mathcal{B}=3.9\times10^{-26}$).
Similarly, Fig. \ref{exampleSchumann} gives parameter estimation results for the Schumann injection.
Interestingly, the distance posterior for this injection \textit{is} consistent with the true Hanford-Livingston separation.
The rotation angle posteriors, though, again exclude the true detector orientations, yielding $\ln\mathcal{B} = -62.7$ ($\mathcal{B}=5.9\times10^{-28}$).

\section{Complications}
\label{complications}

We demonstrated in Sect. \ref{demonstration} above that gravitational-wave geodesy can be successfully used to discriminate between a true stochastic gravitational-wave background and non-astrophysical, terrestrial sources of correlation.
Here, we highlight two important assumptions that have been made in our analysis and discuss what to do should these assumptions not hold.

\subsection{Non-Power-Law Energy-Density Spectra}
\label{complicationsA}

In the main text, we have assumed that our model energy-density spectrum (a power law) is a good description of the true stochastic background.
This assumption was guaranteed by design, as our injected stochastic energy-density spectrum \textit{was} a power-law.
While most gravitational-wave sources are predicted to yield power-law energy-density spectra in the Advanced LIGO and Virgo band, there do exist speculative like superradiant axion clouds \citep{BritoA,BritoB} that may instead yield more complex spectra.

\begin{figure*}
\centering
\includegraphics[width=0.45\textwidth]{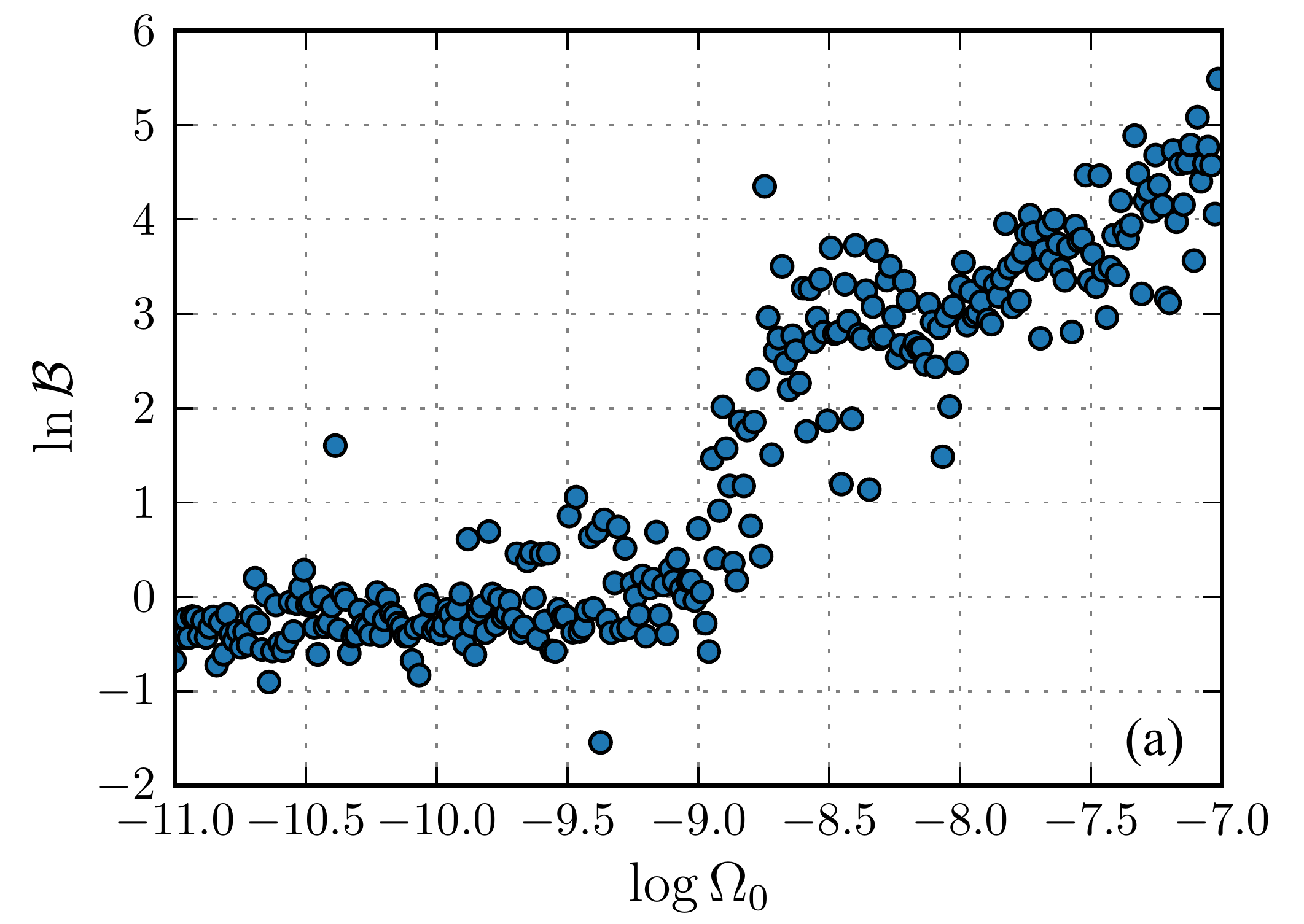}
\includegraphics[width=0.45\textwidth]{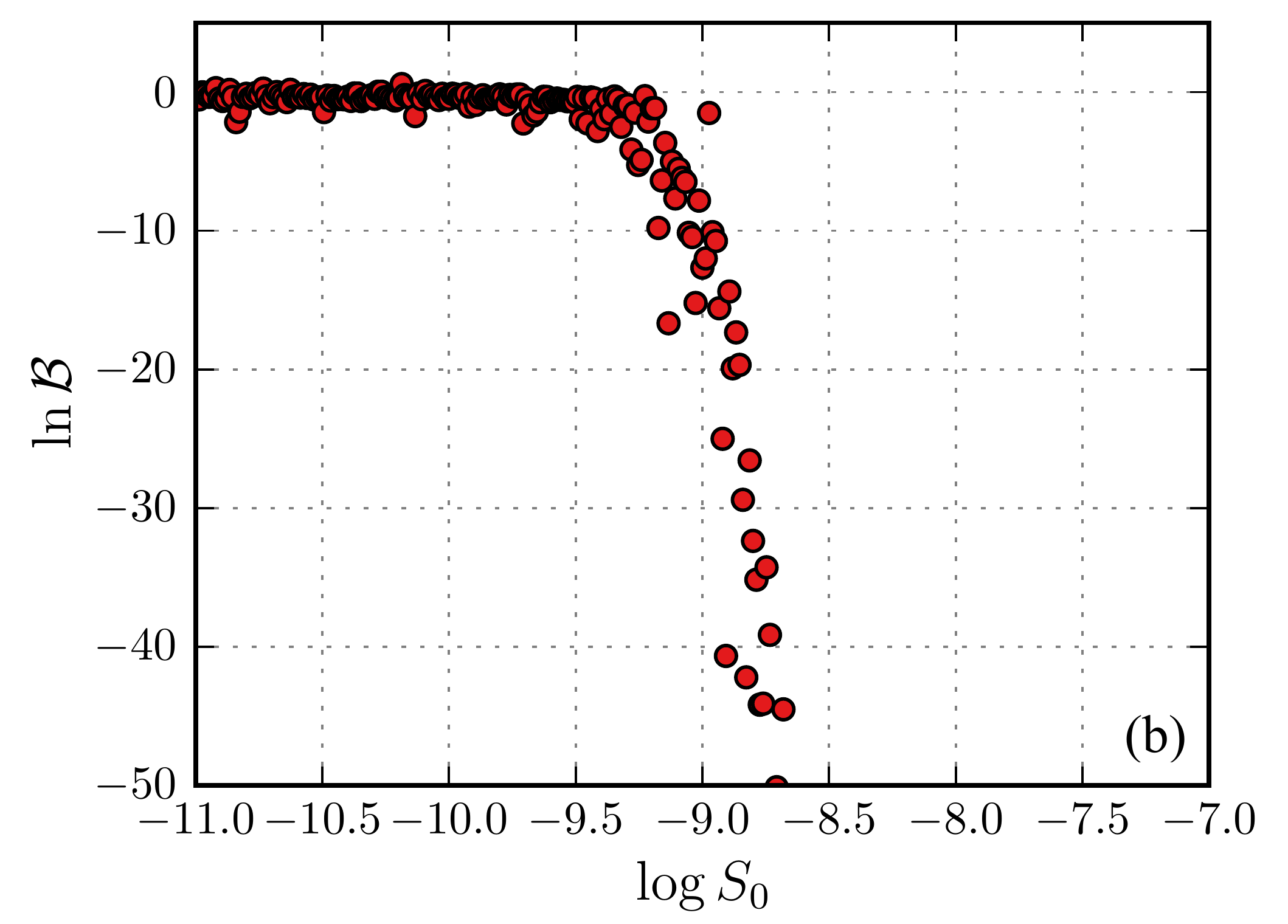}
\caption{Log-Bayes factors between hypotheses $\mathcal{H}_\gamma$ and $\mathcal{H}_\mathrm{Free}$ obtained when analyzing simulated astrophysical broken power-law signals (a) and magnetic Schumann correlations (b), assuming a broken power-law for our model energy-density spectrum.
For each set of simulations, we assume three years of observation with design-sensitivity Advanced LIGO.
When adopting this more complex model, our Bayes factors still scale as expected, with the astrophysical signal preferring $\mathcal{H}_\gamma$ and the Schumann correlations preferring $\mathcal{H}_\mathrm{Free}$.
}
\label{brokenPL}
\end{figure*}

It is worthwhile to investigate how our method fares given more complex energy-density spectra.
Specifically, we will consider observations of a broken power-law background with energy density
	\begin{equation}
	\Omega(f) = \begin{cases}
		\Omega_0 \left(f/f_0\right)^{\alpha_1} & {f\leq f_0} \\
		\Omega_0 \left(f/f_0\right)^{\alpha_2} & {f > f_0}
		\end{cases}.
	\end{equation}
Correspondingly, we will adopt a broken power-law model for $\Omega(f)$ (with free parameters $\Omega_0$, $f_0$, $\alpha_1$ and $\alpha_2$) in both hypotheses $\mathcal{H}_\gamma$ and $\mathcal{H}_\mathrm{Free}$.
We simulate broken power-law signals with slopes $\alpha_1 = 1$, $\alpha_2 = -1$, a knee frequency $f_0=30$ Hz, and amplitudes $\Omega_0$ ranging from $10^{-11}$ to $10^{-7}$.
The recovered Bayes factors between $\mathcal{H}_\gamma$ and $\mathcal{H}_\mathrm{Free}$ are shown in Fig. \ref{brokenPL}a.
We see that, even given a more complex signal and model, our method remains effective.

With a more complex model, is it also true that we can still correctly reject terrestrial sources of correlation?
To verify that the additional free parameters afforded by the broken power-law model do not lead to false acceptances, we again apply the broken-power law model to Schumann correlations of various strengths.
As shown in Fig. \ref{brokenPL}b, the resulting Bayes factors behave as expected, indicating inconsistency with Advanced LIGO's correct geometry.

We have shown that geodesy is a successful discriminator between astrophysical and terrestrial correlations, even when using model more complicated than simple power spectra.
Crucially, though, we have still assumed the \textit{correct} energy-density spectrum, using the same model (a broken power law) to both inject and recover simulated signals.
The most troubling case is the possibility of an \textit{incorrect} model -- one that is a poor descriptor of the true stochastic background.
In this case, would we risk rejecting a real stochastic background as a terrestrial signal?

\begin{figure*}
\centering
\includegraphics[width=0.45\textwidth]{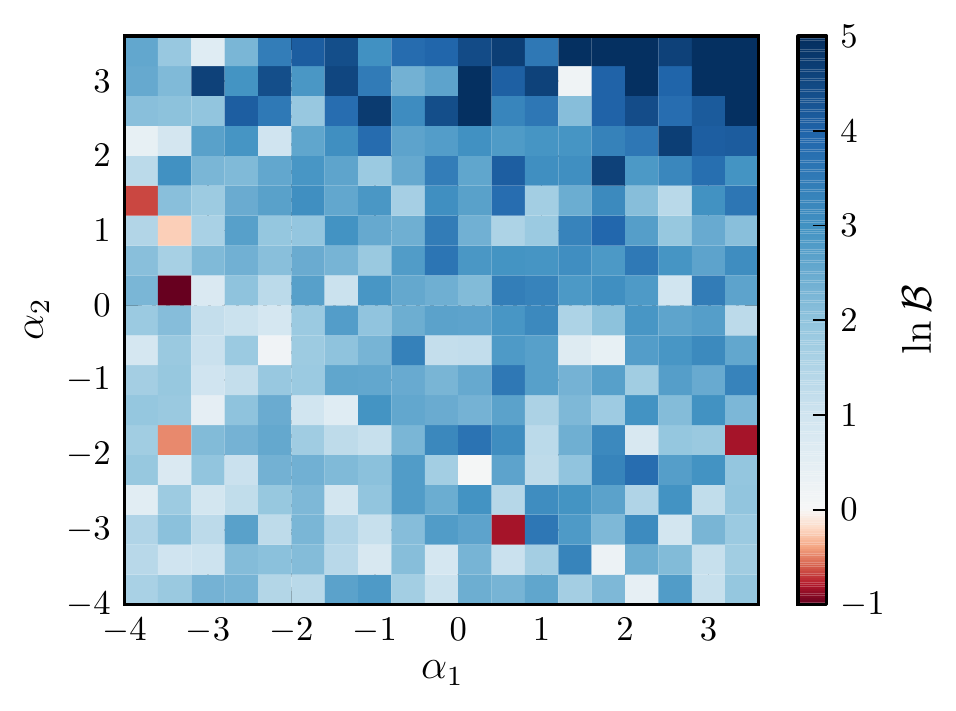}
\caption{Log-Bayes factors between $\mathcal{H}_\gamma$ and $\mathcal{H}_\mathrm{Free}$ when deliberately analyzing astrophysical broken power-law signals with an \textit{incorrect} power-law model.
Each injected signal has a knee frequency of $f_0=30$ Hz and an amplitude $\Omega_0$ scaled such that the signal has $\mathrm{SNR}=5$ after three years of observation with design-sensitivity Advanced LIGO.
Despite the signal-model mismatch, we correctly classify the majority of the simulated signals, with no evidence of increased false-dismissals due to the mismatch.}
\label{varyingSlopes}
\end{figure*}

To test this, we again simulate observations of a broken power-law background, but recover them using an ordinary power-law model, deliberately choosing an incorrect description of the simulated signal.
Figure \ref{varyingSlopes} illustrates the resulting Bayes factors for simulated observations with $\alpha_1$ and $\alpha_2$ each ranging between $-4$ and $4$.
For each injection we chose $f_0=30$ Hz, placing the broken power-law's ``knee'' in the center of the stochastic sensitivity band, and scaled the amplitudes $\Omega_0$ such that each observation has $\mathrm{SNR}=5$ when naively recovered with an ordinary power law.
The vast majority of these simulations yield positive log-Bayes factors, correctly classifying these signals despite our poor choice of model.
Note that the injections falling along the line $\alpha_1 = \alpha_2$ \textit{are} power laws.
If the signal-model mismatch significantly degraded our ability to classify stochastic signals, then Fig. \ref{varyingSlopes} would exhibit a color gradient as we move perpendicularly off the $\alpha_1 = \alpha_2$ line, away from power laws and towards increasingly sharp signal spectra.
Instead, Fig. \ref{varyingSlopes} shows no such gradient, and our method remains robust even in the case of poorly-fitting models.

We attribute this robustness to the fact that the isotropic energy-density spectrum and baseline geometry have very different effects on the expected cross-correlation spectrum $\langle C(f) \rangle = \gamma(f)\Omega(f)$.
The energy density spectrum $\Omega(f)$ is everywhere positive, and so different energy-density spectra can change only the \textit{amplitude} of $C(f)$, not its sign.
The sign of $C(f)$ is set by the overlap reduction function, which alternates between positive and negative values with zero-crossings fixed by the baseline geometry.
Even if our model for $C(f)$ assumes an incorrect energy-density spectrum (as above), our $\mathcal{H}_\gamma$ hypothesis nevertheless predict the correct zero-crossings of the observed cross-correlation spectrum.
This offers some robustness against false-dismissal of a true stochastic signal, even if our model energy-density spectrum is imperfect.
At the same time, it prevents us from \textit{over-fitting} spurious terrestrial correlations [whose sign is unrelated to the sign of $\gamma(f)$], mitigating the risk of false-positives.


\subsection{Anisotropy}
\label{complicationsB}

Second, we have assumed that the stochastic gravitational-wave background is isotropic, giving rise to a cross-correlated signal described by the standard overlap reduction function [Eq. \eqref{orf}].
This is unlikely to be strictly true.
The Solar System's motion with respect to the cosmic microwave background will likely impart a small apparent dipole moment to the stochastic gravitational-wave background.
Additional anisotropies might arise from structure in the local Universe \citep{Jenkins2018,Cusin2018}, as well as the fact that, over a finite integration time, we observe only a discrete set of gravitational-wave events \citep{Meacher2014}.
An anisotropic stochastic background, in contrast, would yield correlations that are \textit{not} consistent with the standard overlap reduction function, but instead with some different effective overlap reduction function.
If we naively analyzed an anisotropic stochastic signal with the method presented in the main text, we would likely find a preference for the (unphysical) hypothesis $\mathcal{H}_\mathrm{Free}$ over $\mathcal{H}_\gamma$ and risk rejecting the signal as terrestrial.

In practice, any anisotropies are unlikely to significantly affect our analysis.
First, expected anisotropies are small.
The Solar System moves with speed $v_\oplus \approx 370$ km/s with respect to the cosmic microwave background, and so the stochastic background's apparent dipole moment is a factor of $v_\oplus/c \sim 10^{-3}$ smaller than the isotropic monopole moment.
True astrophysical anisotropies are also expected to be small.
Considering multipole moments up to $l=20$ (the approximate angular resolution limit of the LIGO Hanford-Livingston baseline; \cite{Thrane2009}), the observed energy density is expected to vary by no more than $\sim10\%$ with direction \citep{Cusin2018,Jenkins2018}.
Second, if anisotropy were a significant concern, the formalism of Sect. \ref{modelSelection} could be straightforwardly extended to accommodate possible anisotropy.

When allowing for anisotropy, the observed energy-density of the stochastic background will generically have directional dependence on our viewing angle $\hatbf n$.
It is generally assumed that an anisotropic energy-density spectrum can be factored via $\Omega(\hatbf n,f) = H(f) \mathcal{P}(\hatbf n)$, where $H(f)$ and $\mathcal{P}(\hatbf n)$ encode the frequency and directional dependence of $\Omega(\hatbf n, f)$, respectively.
We can further decompose $\mathcal{P}(\hatbf n)$ into a sum of spherical harmonics $Y_{lm}(\hatbf n)$, giving \citep{Allen1997,Thrane2009,DirectionalO1}
	\begin{equation}
	\Omega(\hatbf n,f) = H(f) \sum_{l,m} \mathcal{P}^{lm} Y_{lm}(\hatbf n)
	\end{equation}
for some set of coefficients $\mathcal{P}^{lm}$.
We use the normalization convention $\int | Y_{lm}(\hatbf n) |^2 \dd\hatbf{n} = 1$.

Over the course of a sidereal day, gravitational-wave detectors have varying sensitivities to different sky directions $\hatbf n$.
In the presence of an anisotropic background, the expected cross-correlation between detectors is therefore time-dependent:
	\begin{equation}
	\label{anisotropicC}
	\langle C(f,t) \rangle = H(f) \sum_{l,m} \mathcal{P}^{lm}\gamma_{lm}(t,f),
	\end{equation}
where $t$ is periodic over a sidereal day.
This expression is similar in form to Eq. \eqref{avg}, but with a sum over spherical harmonics and distinct (time-dependent) overlap reduction functions for each spherical harmonic \citep{Allen1997,Thrane2009}
	\begin{equation}
	\label{anisotropicOrf}
	\begin{aligned}
	\gamma_{lm}(t,f) = \frac{5}{2\sqrt{4\pi}} \sum_A \int_\mathrm{Sky}
		&Y_{lm}(\hatbf n) F^A_1(\hatbf n,t) F^A_2(\hatbf n,t) \\
		&\times \mathrm{e}^{2\pi i f \Delta\mathbf{x}(t)\cdot\hatbf{n}/c}
		 \dd\hatbf{n}.
	\end{aligned}
	\end{equation}
In Eq. \eqref{anisotropicOrf}, the detectors' antenna patterns $F^A_{i}(\hatbf n, t)$ and separation vector $\Delta \mathbf{x}(t)$ are time-dependent, rotating with the Earth over the course of a sidereal day.
The normalization of Eq. \eqref{anisotropicOrf} is chosen such that monopole overlap reduction function $\gamma_{00}(t,f)$ reduces to Eq. \eqref{orf} above.
The time-dependence of Eq. \eqref{anisotropicOrf} can be conveniently factored out via \citep{Allen1997,Thrane2009}
	\begin{equation}
	\gamma_{lm}(t,f) = \gamma_{lm}(0,f) \mathrm{e}^{2\pi i m (t/T) },
	\end{equation}
where $T$ is the length of one sidereal day.

If we incorrectly assumed an isotropic background and averaged our cross-correlation measurements over a sidereal day, we would measure cross-correlation
	\begin{equation}
	\label{integratedAnisC}
	\begin{aligned}
	\langle C(f) \rangle &= \frac{1}{T}\int_0^T \langle C(f,t)\rangle dt \\
		&= H(f) \sum_{l,m} \mathcal{P}^{lm} \gamma_{lm}(0,f) \frac{1}{T} \int_0^T \mathrm{e}^{2\pi i m (t/T) }dt \\
		&= H(f) \sum_l \mathcal{P}^{l0} \gamma_{l0}(0,f),
	\end{aligned}
	\end{equation}
where the integral vanishes for all $m\ne 0$.
Equation \eqref{integratedAnisC} does not trace the isotropic overlap reduction function, but instead follows a linear combination of the anisotropic $\gamma_{l0}(f)$'s.
Thus, if the background were significantly anisotropic (with some $\mathcal{P}^{l0}$ comparable in magnitude to the monopole amplitude $\mathcal{P}^{00}$), we would incorrectly conclude that the resulting correlated signal is incompatible with our detector geometry and dismiss it as terrestrial.

In analogy to Eq. \eqref{GammaC}, one could define hypothesis $\mathcal{H}_\gamma$ via the model
	\begin{equation}
	C_\gamma(\Theta, \mathcal{P}^{lm};f) = H(\Theta;f) \sum_{l,m} \mathcal{P}^{lm} \gamma^\mathrm{True}_{lm}(t,f),
	\end{equation}
where $\gamma^\mathrm{True}_{lm}(t,f)$ is the baseline's known overlap reduction function for spherical harmonic $(l,m)$ and $\Theta$ represents the variables parametrizing $H(f)$.
Similarly, the unphysical hypothesis $\mathcal{H}_\mathrm{Free}$ would become
	\begin{equation}
	C_\mathrm{Free}(\Theta,\mathcal{P}^{lm},\theta,\phi_1,\phi_2;f) 
		= H(\Theta;f) \sum_{l,m}  \mathcal{P}^{lm} \gamma_{lm}(\theta,\phi_1,\phi_2;t,f).
	\end{equation}

\end{document}